\pdfoutput=1
\RequirePackage{ifpdf}
\ifpdf 
\documentclass[pdftex]{sigma}
\else
\documentclass{sigma}
\fi

\numberwithin{equation}{section}

\newtheorem{Theorem}{Theorem}[section]
\newtheorem{Proposition}[Theorem]{Proposition}
\newtheorem{Conjecture}[Theorem]{Conjecture}
 { \theoremstyle{definition}

\newtheorem{Remark}[Theorem]{Remark} }

\begin{document}


\renewcommand{\thefootnote}{$\star$}

\newcommand{\arXivNumber}{1607.01351}

\renewcommand{\PaperNumber}{105}

\FirstPageHeading

\ShortArticleName{On the Tracy--Widom$_\beta$ Distribution for $\beta=6$}

\ArticleName{On the Tracy--Widom$\boldsymbol{_\beta}$ Distribution for $\boldsymbol{\beta=6}$\footnote{This paper is a~contribution to the Special Issue on Asymptotics and Universality in Random Matrices, Random Growth Processes, Integrable Systems and Statistical Physics in honor of Percy Deift and Craig Tracy. The full collection is available at \href{http://www.emis.de/journals/SIGMA/Deift-Tracy.html}{http://www.emis.de/journals/SIGMA/Deift-Tracy.html}}}

\Author{Tamara GRAVA~$^{\dag^1\dag^2}$, Alexander ITS~$^{\dag^3}$, Andrei KAPAEV~$^{\dag^4}$ and Francesco MEZZADRI~$^{\dag^1}$}
\AuthorNameForHeading{T.~Grava, A.~Its, A.~Kapaev and F.~Mezzadri}

\Address{$^{\dag^1}$~School of Mathematics, University of Bristol, Bristol, BS8 1SN, UK}
\EmailDD{\href{mailto:tamara.grava@bristol.ac.uk}{tamara.grava@bristol.ac.uk}, \href{mailto:francesco.mezzadri@bristol.ac.uk}{francesco.mezzadri@bristol.ac.uk}}

\Address{$^{\dag^2}$~SISSA, via Bonomea 265, 34100, Trieste, Italy}
\EmailDD{\href{mailto:grava@sissa.it}{grava@sissa.it}}

\Address{$^{\dag^3}$~Department of Mathematical Sciences, Indiana University -- Purdue University Indianapolis,\\
\hphantom{$^{\dag^3}$}~Indianapolis, IN 46202-3216, USA}
\EmailDD{\href{mailto:aits@iupui.edu}{aits@iupui.edu}}

\Address{$^{\dag^4}$~Department of Mathematical Physics, St.~Petersburg State University, St.~Petersburg, Russia}
\EmailDD{\href{mailto:kapaev55@mail.ru}{kapaev55@mail.ru}}

\ArticleDates{Received July 04, 2016, in f\/inal form October 25, 2016; Published online November 01, 2016}

\Abstract{We study the Tracy--Widom distribution function for Dyson's $\beta$-ensemble with $\beta = 6$. The starting point of our analysis is the recent work of I.~Rumanov where he produces a Lax-pair representation for the Bloemendal--Vir\'ag equation. The latter is a linear PDE which describes the Tracy--Widom functions corresponding to general values of $\beta$. Using his Lax pair, Rumanov derives an explicit formula for the Tracy--Widom $\beta=6$ function in terms of the second Painlev\'e transcendent and the solution of an auxiliary ODE. Rumanov also shows that this formula allows him to derive formally the asymptotic expansion of the Tracy--Widom function. Our goal is to make Rumanov's approach and hence the asymptotic analysis it provides rigorous. In this paper, the f\/irst one in a sequel, we show that Rumanov's Lax-pair can be interpreted as a certain gauge transformation of the standard Lax pair for the second Painlev\'e equation. This gauge transformation though contains functional parameters which are def\/ined via some auxiliary nonlinear ODE which is equivalent to the auxiliary ODE of Rumanov's formula. The gauge-interpretation of Rumanov's Lax-pair allows us to highlight the steps of the original Rumanov's method which needs rigorous justif\/ications in order to make the method complete. We provide a rigorous justif\/ication of one of these steps. Namely, we prove that the Painlev\'e function involved in Rumanov's formula is indeed, as it has been suggested by Rumanov, the Hastings--McLeod solution of the second Painlev\'e equation. The key issue which we also discuss and which is still open is the question of integrability of the auxiliary ODE in Rumanov's formula. We note that this question is crucial for the rigorous asymptotic analysis of the Tracy--Widom function. We also notice that our work is a partial answer to one of the problems related to the $\beta$-ensembles formulated by Percy Deift during the June 2015 Montreal Conference on integrable systems.}


\Classification{30E20; 60B20; 34M50}

\rightline{\it Dedicated to Percy Deift and Craig Tracy on the occasion of their 70th birthdays}

\renewcommand{\thefootnote}{\arabic{footnote}}
\setcounter{footnote}{0}

\section{Introduction}

Given $\beta >0$, Dyson's $\beta$-ensemble is def\/ined as a Coulomb gas of $N$ charged particles, that is as the space of $N$ one dimensional particles, $\{-\infty < \lambda_1 <\lambda_2 < \cdots < \lambda_{N} < \infty\}$ with the probability density given by the equation
\begin{gather}\label{densityeigenvalues}
	 p(\lambda_1,\ldots,\lambda_N)d\lambda_1\cdots d\lambda_N = \frac{1}{Z_N} \prod_{1\leq j,k \leq N }|\lambda_j-\lambda_k|^\beta {\rm e}^{-\beta \underset{j=1}{\overset{N}{\sum}}V(\lambda)}d\lambda_1\cdots d\lambda_N,
\\
Z_{N} = \int_{-\infty}^{\infty} \cdots \int_{-\infty}^{\infty}\prod_{1\leq j,k \leq N }|\lambda_j-\lambda_k|^\beta {\rm e}^{-\beta \underset{j=1}{\overset{N}{\sum}}V(\lambda)}d\lambda_1\cdots d\lambda_N.\nonumber
\end{gather}
Here, $V(\lambda)$ has a meaning of external f\/ield which we will assume to be Gaussian, i.e., $V(\lambda) = \frac{\lambda^2}{2}$.
 The objects of interest are the gap probabilities in the large $N$ limit. We will be particularly concerned with the soft edge probability distribution
\begin{gather*}
F_{\beta}(t) \equiv E_{\beta}^{{\rm soft}}\bigl(0; (t, \infty)\bigr) = \lim_{N\rightarrow \infty}E_{\beta N}^{{\rm soft}}\left(0;\left(\sqrt{2N}
+\frac{t}{\sqrt{2}N^{1/6}}, \infty\right)\right),
\end{gather*}
where
\begin{gather}\label{mint}
E_{\beta N}^{{\rm soft}}\bigl(0; (t, \infty)\bigr) = \int_{-\infty}^{t} \cdots \int_{-\infty}^{t} p(\lambda_1,\ldots,\lambda_N)d\lambda_1\cdots d\lambda_N.
\end{gather}
Cases $\beta = 1, 2, 4$ known as Gaussian orthogonal (GOE), Gaussian unitary (GUE) and Gaussian symplectic (GSE) ensembles. Indeed, in these cases, distribution~(\ref{densityeigenvalues}) describes the statistics of the eigenvalues of orthogonal, Hermitian, and symplectic random matrices, respectively, with i.i.d.\ matrix entries. The corresponding limiting edge distribution functions $F_{\beta}(t)$ then become the classical {\it Tracy--Widom distributions}~\cite{TW1}. They admit explicit representations as either the Airy kernel Fredholm determinants or in terms of the Hastings--McLeod solution of the second Painlev\'e equation. These representations, in turn, allow one to evaluate the asymptotic expansions of $F_{\beta}(t)$ as $t \to -\infty$.

In this paper we address the question of the asymptotic analysis of~$F_{\beta}(t)$ beyond the classical values $\beta = 1, 2, 4$. The crucial problem is that the orthogonal polynomial approach, which is the principal technique in the intergrable random matrix case, is not available for general~$\beta$. However, several highly nontrivial conjectures concerning the general $\beta$ ensembles have been suggested. An excellent presentation of the state of art in this area is given in the survey by P.~Forrester~\cite{For}. The current principal heuristic result concerning the asymptotic behavior of the generalized Tracy--Widom distribution $F_{\beta}(t)$ was obtained in 2010 by G.~Borot, B.~Eynard, S.N.~Majumdar and C.~Nadal and it reads as follows.
\begin{Conjecture}[\cite{BEMN}] \label{bemncon1}
\begin{gather}
F_{\beta}(t) = \exp\left(-\beta\frac{|t|^3}{24}
+ \frac{\sqrt{2}}{3} (\beta/2 - 1 )|t|^{3/2}\right.\nonumber\\
\left.\hphantom{F_{\beta}(t) =}{} + \frac{1}{8} (\beta/2 + 2/\beta - 3 )\log|t|
+ c_0 + O\left(\frac{1}{|t|^{3/2}}\right)\right), \qquad t \to -\infty.\label{con1}
\end{gather}
The constant term $c_0$ is also explicitly predicted. Indeed, it is claimed that
\begin{gather}
c_0 = \frac{\beta}{2}\left(\frac{1}{12} - \zeta'(-1) \right) + \frac{\gamma}{6\beta} -\frac{\log2\pi}{4} - \frac{(\beta/2)}{2}
 \nonumber\\
\hphantom{c_0 =}{}
 +\left(\frac{17}{8} - \frac{25}{24} (\beta/2 + 2/\beta )\right)\log 2
 +\int_{0}^{\infty}\frac{1}{e^{\beta t/2} - 1} \left(\frac{t}{e^t - 1} -1 +\frac{t}{2} -\frac{t^2}{12}\right)dt,\label{conconst}
 \end{gather}
 where $\zeta(z)$ is Riemann's zeta-function and $\gamma$ denotes Euler's constant.
 \end{Conjecture}

Formulae (\ref{con1}),~(\ref{conconst}) have been derived in \cite{BEMN} within the framework of the so-called loop-equation technique by performing the relevant double scaling limit directly in the formal large~$N$ expansion of the multiple integral in~(\ref{mint}).

\begin{Remark} For the classical cases $\beta =1, 2, 4$, formula (\ref{con1}), without the constant term, was obtained and proved in \cite{TW1} and \cite{TW2} by using the (established in these papers) representations of~$F_{\beta =1,2,4}(t)$ in terms of the Hastings--McLeod solution of the second Painlev\'e equation. Moreover, in \cite{TW1,TW2} the value (\ref{conconst}) of the constant $c_0$ for $\beta = 1, 2, 4$ was also conjectured. Rigorous derivation of~(\ref{con1}),~(\ref{conconst}) for the case $\beta =2$ is given in \cite{BBDiF,DIK}. In~\cite{BBDiF}, the cases $\beta =1, 4$ are also done. The papers \cite{DIK} and \cite{BBDiF} employ the Riemann--Hilbert approach and the Deift--Zhou nonlinear steepest descent method~\cite{DZ} which are available to the classical cases of $\beta = 1, 2, 4$.
\end{Remark}

\begin{Remark} The leading asymptotic term in (\ref{con1}) for arbitrary $\beta$ has been rigorously obtained in~\cite{RRV} with the help of the analysis of the certain stochastic Schr\"odinger operator. We shall mention this paper again right after the next remark.
\end{Remark}

\begin{Remark} For the limiting hard edge probability distribution, the heuristic asymptotic result similar to~(\ref{con1}) was obtained in~\cite{CM}. In fact, in the case of positive integer values of the exponent in the corresponding Lauguerre weight the asymptotic formula had been already proven, including the rigorous derivation of the constant term, in~\cite{For2}. For more on the hard edge case we refer the interested reader again to survey~\cite{For}.
\end{Remark}

It is remarkable, that paper \cite{BEMN}, while giving such detailed formulae for the asymptotics of the distribution function~$F_{\beta}(t)$ does not actually produce any description of the object itself (for f\/inite values of~$t$). The latter has been done by A.~Bloemendal and B.~Vir\'ag~\cite{BV}. Inspired by the pioneering work of E.~Dumitriu and A.~Edelman~\cite{DE} and by the subsequent works~\cite{VV} and~\cite{RRV}, Bloemendal and Vir\'ag~\cite{BV} connect the analysis of the generalized Tracy--Widom distribution~$F_{\beta}(s)$ to the study of stochastic Schr\"odinger operators. In particular, it has been proven in~\cite{BV} that the Tracy--Widom distribution function $F_{\beta}(t)$, for any $\beta$, can be expressed in terms of the solution of a certain linear PDE. In more details, the result of~\cite{BV} can be formulated as follows.

Consider the partial dif\/ferential equation for the scalar function $F(x,t;\beta)$
\begin{gather}\label{bv1}
\frac{\partial F}{\partial t} +\frac{2}{\beta}\frac{\partial^2F}{\partial x^2} + \big(t - x^2\big)\frac{\partial F}{\partial x} = 0,\qquad (x, t)\in {\mathbb R}^2
\end{gather}
supplemented by the boundary conditions
\begin{gather}\label{bv2}
F(x, t;\beta) \to 1, \qquad \mbox{as}\quad x, t \to \infty,
\end{gather}
and
\begin{gather}\label{bv3}
F(x, t;\beta) \to 0, \qquad \mbox{as}\quad x \to - \infty \quad \mbox{for f\/ixed $t$.}
\end{gather}

\begin{Theorem}[\cite{BV}]\label{th1}
The boundary value problem \eqref{bv1}--\eqref{bv3} has a unique bounded smooth solution. Moreover, equation
\begin{gather*}
F_{\beta} (t) = \lim_{x \to \infty} F(x, t;\beta),
\end{gather*}
determines the Tracy--Widom distribution function for the general value of the parameter $\beta >0$.
\end{Theorem}

Equation (\ref{bv1}) has a very interesting interpretation from the point of view of the theory of Painlev\'e transcendents. In fact, there are two ways to connect (\ref{bv1}) with the Painlev\'e equations. The f\/irst way identif\/ies~(\ref{bv1}) with the {\it quantum} second Painlev\'e equation. Indeed, the second Painev\'e equation, i.e., the equation
\begin{gather}\label{p2int}
u_{tt} = tu + 2u^3,
\end{gather}
admits the Hamiltonian form \cite{Okamoto}
\begin{gather*}
u_t = \frac{\partial H}{\partial p}, \qquad p_t = -\frac{\partial H}{\partial u},
\end{gather*}
where the (time dependent) Hamiltonian $H\equiv H(t, u, p)$ is
\begin{gather*}
H = \frac{p^2}{2} -\left(u^2 +\frac{t}{2}\right)p -\frac{u}{2}.
\end{gather*}
Set
\begin{gather*}
\Phi(u,t;\beta) := F\bigl({-}2^{1/3}e^{-i\frac{\pi}{3}}u, 2^{-1/3}e^{i\frac{\pi}{3}}t;\beta\bigr).
\end{gather*}
Then, in terms of $\Phi(u,t)$, equation (\ref{bv1}) becomes the imaginary time Schr\"{o}dinger equation generated by the quantum Hamiltonian $H(t,u,\hbar\partial_u) +u/2$ with $2/\beta$ playing the role of Planck's constant $\hbar$. That is, we have that
\begin{gather*}
\frac{2}{\beta}\partial_t\Phi = \left(H\left(t , u, \frac{2}{\beta}\partial_u\right) +\frac{u}{2}\right)\Phi.
\end{gather*}
\begin{Remark} Equation (\ref{bv1}) in its original variables can be also interpreted as an imaginary time Schr\"{o}dinger equation generated by the quantum Hamiltonian $H_0(t, x, \hbar\partial_x)$ with
\begin{gather*}
H_0(t, x, p) := -p^2 - \big(t - x^2\big)p
\end{gather*}
and with the Planck constant $\hbar$ again equals $2/\beta$. The corresponding classical dynamical system is equivalent to the Painlev\'e 34 equation for the momentum~$p(t)$
\begin{gather*}
p_{tt} = 4p^2 +2pt + \frac{p^2_t}{2p},
\end{gather*}
which, in turn, is reduced again to the second Painlev\'e equation (\ref{p2int}) for the function $u = \sqrt{p}$.
\end{Remark}

We refer the reader to the papers \cite{Nagoya,Rumanov2,Slavyanov,Suleimanov,ZZ} for more on the quantum Painlev\'e equations and their connections to the $\beta$-ensembles. It should be also mentioned that in the current state of development of this subject, this connection provides us with no ef\/f\/icient tools of asymptotic analysis of the $\beta$-ensembles.

The second connection of equation (\ref{bv1}) to the Painlev\'e theory has been already found in~\cite{BV}, and it gives indeed an ef\/f\/icient apparatus for the analysis of (\ref{bv1}), but, at the moment, only for the classical value $\beta =2$. As it is shown in~\cite{BV}, if $\beta =2$, the solution of (\ref{bv1}) is reduced to the solution of the Riemann--Hilbert problem canonically associated with the Painlev\'e~II equation~(\ref{p2int}). More precisely, this important observation can be described as follows.

The second Painlev\'e equation (\ref{p2int}) is the compatibility condition of the following Flaschka--Newell \cite{FN} Lax pair
\begin{gather}\label{laxint1}
\dfrac{d\Psi_0}{dx}=\hat{L}_0\Psi_0,\qquad \dfrac{d\Psi_0}{dt}=\hat{B}_0\Psi_0,\quad
\end{gather}
where the matrices $\hat{L}_0$ and $\hat{B}_0$ are
\begin{gather*}
\hat{L}_0=\frac{x^2}{2}\sigma_3+x\begin{pmatrix}0&u\\ u&0\end{pmatrix}+\begin{pmatrix}-\dfrac{t}{2} -u^2&-u_t\\u_t&\dfrac{t}{2} + u^2\end{pmatrix}
\end{gather*}
and
\begin{gather*}
\hat{B}_0=-\frac{x}{2}\sigma_3-\begin{pmatrix}0&u\\ u&0\end{pmatrix},
\end{gather*}
where $\sigma_3=\begin{pmatrix}1&0\\0&-1\end{pmatrix}$. Put
\begin{gather}\label{laxint4}
F(x,t;\beta=2) = (\Psi_0(x,t))_{22}e^{\frac{x^3}{6} - \frac{1}{2}tx +\int^{\infty}_{t}\omega(\tau)d\tau}, \qquad \omega = u^4 +tu^2 - u^2_t.
\end{gather}
Then $F(x,t;\beta=2)$ will satisfy (\ref{bv1}) with $\beta=2$. It is also shown in \cite{BV} that the boundary conditions (\ref{bv2}), (\ref{bv3}) select in formula (\ref{laxint4}) the Hastings--McLeod \cite{HM} solution of~(\ref{p2int}) uniquely def\/ined by the asymptotic condition
\begin{gather*}
u(t) \simeq \operatorname{Ai}(t), \qquad t \to +\infty,
\end{gather*}
and having the following behavior on the other end of the real line
\begin{gather*}
u(t) \sim \sqrt{-\dfrac{t}{2}},\qquad t \to -\infty.
\end{gather*}
Here, $\operatorname{Ai}(x)$ denotes the usual Airy function. This fact yields immediately the Tracy--Widom formula for $F_{2}(t)$,
\begin{gather*}
F_2(t) = e ^{\int^{\infty}_{t}\omega(\tau)d\tau},
\end{gather*}
and, with the reference to \cite{BBDiF,DIK}, the proof of Conjecture~\ref{bemncon1} above for $\beta = 2$.

It is tempting to f\/ind the analog of the Lax pair representation~(\ref{laxint4}) for the solution of~(\ref{bv1}) for arbitrary $\beta$. For the case of even values of $\beta$, important progress toward this goal has been achieved by I.~Rumanov in \cite{Rumanov0,Rumanov}. In particular, in~\cite{Rumanov} Rumanov has produced a formula very similar to (\ref{laxint4}) for the f\/irst interesting case, $\beta=6$. However, Rumanov's higher beta analogue of the Lax pair (\ref{laxint1}) involves functional parameters which are def\/ined via auxiliary nonlinear ODEs, and this makes Rumanov's approach for $\beta > 4$ less ef\/f\/icient than in the classical cases of $\beta = 1, 2, 4$. Moreover, several important steps in Rumanov's approach are only justif\/ied on a~heuristic level. In this manuscript we are suggesting a~new interpretation of the results of \cite{Rumanov} which allows us to f\/ill some of the gaps of Rumanov's scheme.

Our f\/irst result is to show that Rumanov's Lax pair in the case $\beta=6$ can be obtained by a gauge transformation of the Painlev\'e II Lax pair.
Let $\Psi_0(x,t)$ be the fundamental solution of the Painlev\'e~II Lax pair (\ref{laxint1}) and let us introduce the matrix function $\Psi(x,t)$ def\/ined as
\begin{gather}\label{bgauge01}
\Psi(x,t) := e^{\frac{x^3}{6} -\frac{xt}{2}}\kappa(t)\begin{pmatrix}
\dfrac{1+q_2(t)}{2}x-\alpha(t)&-1\vspace{1mm}\\
\dfrac{1-q_2^2(t)}{4}&0\end{pmatrix}e^{-\frac{i\pi}{2}\sigma_3}
\left(\frac{1}{u(t)}\right)^{\frac{\sigma_3}{2}}\Psi_0(x,t),
\end{gather}
where $\kappa(t)$, $q_2(t)$ and $\alpha(t)$ are free functional parameters and $u(t)$ is the solution of the PII equation~(\ref{p2int}). Then our key observation is that the matrix function $\Psi(x,t)$ satisf\/ies a Lax pair equivalent to Rumanov's Lax pair. Furthermore, using some of the constructions of~\cite{Rumanov} presented in Section~\ref{rlax} of our paper, we arrive in Section~\ref{gauge} at the following conclusion. We show that the equation
\begin{gather}\label{F6}
F(x,t;\beta=6) = \Psi_{11}\bigl(3^{1/3}x, 3^{2/3}t\bigr),
\end{gather}
where the function $\Psi(x,t)$ is given by (\ref{bgauge01}), determines a solution of the Bloemendal--Vir\'ag equation~(\ref{bv1}) for $\beta =6$ if in addition one has
\begin{gather}\label{kappaint}
\frac{\kappa_t}{\kappa} = -\frac{1}{3}\omega -\frac{2}{3}\alpha -\frac{u_t}{u}\frac{1-2q_2}{6},
\end{gather}
and demands that the functions $q_2(t)$ and $\alpha(t)$ satisfy the following nonlinear ODEs
\begin{gather}\label{q2dif0}
q_{2t} = q_2\left(\frac{2}{3}\alpha + \frac{u_t}{u}\frac{2-q_2}{3}\right) +\frac{u_t}{u}\frac{2-q_2}{3},
\end{gather}
and
\begin{gather}\label{alpha2dif0}
\alpha_{t} = \alpha\left(\frac{2}{3}\alpha + \frac{u_t}{u}\frac{2-q_2}{3}\right) - \frac{t}{6}(1+q_2)-
\frac{u^2}{3}(3+q_2).
\end{gather}
Equations (\ref{kappaint})--(\ref{alpha2dif0}) are equivalent to the above mentioned auxiliary nonlinear ODEs of Rumanov. The next question is how to ref\/lect in this construction the boundary conditions (\ref{bv2}), (\ref{bv3}).

\begin{Conjecture}[\cite{Rumanov}]\label{rumcon} Equation \eqref{F6} determines the solution of the boundary value problem \eqref{bv1}--\eqref{bv3} if the Painlev\'e function $u(t)$ is the Hastings--McLeod solution of \eqref{p2int} and the pair $(q_2(t), \alpha(t))$ is the solution of the system \eqref{q2dif0}--\eqref{alpha2dif0} satisfying the following initial conditions at $t =+\infty$
\begin{gather}\label{assum0int}
q_2 = -1 + o(1), \qquad \alpha = o(1) , \qquad t \to +\infty.
\end{gather}
In addition, the function $\kappa$ and the branch $u^{1/2}$ in \eqref{bgauge01} should be f\/ixed so that
\begin{gather*}
\kappa u^{\frac{1}{2}} \to 1, \qquad \mbox{as}\quad t \to +\infty.
\end{gather*}
\end{Conjecture}

This conjecture should be supplemented by yet another conjecture concerning the system (\ref{q2dif0}), (\ref{alpha2dif0}).
\begin{Conjecture}\label{rumcon2} The system \eqref{q2dif0}, \eqref{alpha2dif0} has a unique smooth solution $(q_2, \alpha)$ that satisfy conditions~\eqref{assum0int} as $t \to + \infty$.
\end{Conjecture}

Assuming that Conjectures \ref{rumcon} and \ref{rumcon2} are true, the Tracy--Widom distribution function for $\beta = 6$ admits the following representation in terms of the Hastings--McLeod Painlev\'e func\-tion~$u(t)$ and the auxiliary function $q_2(t)$
 \begin{gather}\label{TWanswer20}
 F_{6}\big(3^{-2/3}t\big) = \frac{(q_2-1)}{2q_2} \exp\left(\frac{1}{3}\int_{t}^{\infty}\omega(s)ds -\frac{2}{3}\int_{t}^{\infty}\frac{u_s(s)}{u(s)}\frac{1+q_2(s)}{q_2(s)}ds\right),
 \end{gather}
 with $\omega$ as in (\ref{laxint4}). Although not identical, this formula is equivalent to Rumanov's formula~(1.13) in \cite{Rumanov}.

The necessity to analyze the additional non-trivial dif\/ferential equations, i.e., equations (\ref{q2dif0}), (\ref{alpha2dif0}), makes representation (\ref{TWanswer20}) not quite good for an ef\/fective analysis. However, we believe that this is an important step toward the rigorous theory of the $\beta$-ensembles with the general value of~$\beta$. In fact, as it is shown in~\cite{Rumanov}, one can derive from (\ref{q2dif0}), (\ref{alpha2dif0}), at least formally, a power series expansion of $q_2(t)$ as $t\to -\infty$ which, when substituted into~(\ref{TWanswer20}), would match, except for the constant term~$\chi$, the asymptotic formula for $F_{6}(t)$ from Conjecture~\ref{bemncon1}. We reproduce this result in Section~\ref{assymp}.

This paper is the f\/irst in a series where we intend to transform Rumanov's approach into a rigorous scheme. The main goal of this paper is to prove Conjecture~\ref{rumcon}, assuming that Conjecture~\ref{rumcon2} is true. Our proof of Conjecture~\ref{rumcon} is based on the already described observation that Rumanov's Lax pair is gauge equivalent by the transformation~(\ref{bgauge01}) to the standard Lax pair~(\ref{laxint1}) for the second Painlev\'e equation. The auxiliary functions $q_2(t)$, $\alpha(t)$ and $\kappa(t)$ appear as functional parameters of this gauge transformation. We expect that this gauge equivalence to the Painlev\'e II Lax pair takes place for all Lax pairs which are found by Rumanov for even~$\beta$.

\begin{Remark} The functional parameters used in \cite{Rumanov} are denoted as $q_2(t)$, $q_1(t)$, $q_0(t)$ and $U(t)$.
The relations to the parameters $q_2(t)$, $\alpha(t)$, $u(t)$ and $\kappa(t)$ which we use here are given by the equations
\begin{gather*}
q_2 = q_2, \qquad q_1 = 2\alpha +\frac{u_t}{u}(1 + q_2),\qquad q_0 = 2\alpha\frac{u_t}{u} + t + 2u^2,\\
U(t) = 6\frac{d}{dt}\left(\log\frac{\kappa}{\sqrt{1-q^2_2}}\right) -\frac{t^2}{2}.
\end{gather*}
\end{Remark}

\begin{Remark} We expect that the auxiliary ODEs (\ref{q2dif0}) and (\ref{alpha2dif0}) can be put within the context of integrable systems and this will complete the analysis of the Tracy--Widom distribution function for $\beta =6$. We shall discuss this and other open issues related to Rumanov's approach in the concluding section of this paper where we shall also clarify our use of the term ``integrability''.
\end{Remark}

\section{Rumanov's Lax pair}\label{rlax}
The original Rumanov's Lax pair is the following linear system of two $2\times 2$ matrix dif\/ferential equations
\begin{gather}\label{lax1}
\frac{d\Psi}{dx}=L\Psi,\qquad \frac{d\Psi}{dt}=B\Psi,
\end{gather}
where
\begin{gather}\label{lax2}
L=\frac{1}{2}\begin{pmatrix}x^2-t+x^2q_2-xq_1+q_0&2(x^3-x^2e_1+xe_2-e_3)\\
(x+e_1)\dfrac{1-q_2^2}{2}+q_1q_2&x^2-t-x^2q_2+xq_1-q_0
\end{pmatrix},
\end{gather}
and
\begin{gather}\label{lax3}
B=\begin{pmatrix}
-\dfrac{x}{2}(1+q_2)+{a}&-x^2+xb+c\\
\dfrac{q_2^2-1}{4}&-\dfrac{x}{2}(1-q_2)+{d}
\end{pmatrix},
\end{gather}
where
\begin{gather}\label{atild}
{a}={d}+q_2(b-e_1)+q_1.
\end{gather}
Let us also introduce the functional parameter
\begin{gather}\label{Udef}
U = 3(a+d) -\frac{t^2}{2}.
\end{gather}
We note that
\begin{gather*}
\operatorname{Tr} B=-x+\frac{t^2}{6}+\frac{U}{3}.
\end{gather*}
The compatibility of equations (\ref{lax1}) implies
\begin{gather}\label{lax22}
\frac{d B}{dx}-\frac{d L}{dt}=[L,B],
\end{gather}
which is equivalent to the following set of equations for the parameters $e_1$, $e_2$, $e_3$ and $q_0$, $q_1$ and~$q_2$:
\begin{gather}
\frac{d e_1}{dt}=(b-e_1)(q_2e_1-q_1)+q_2(c+e_2)-q_0,\label{rumsys1}\\
\frac{d e_2}{dt}=-2 +q_2(be_2+e_3-e_1e_2)+q_1e_2+q_1c-q_0b,\\
\frac{d e_3}{dt}=e_3(q_1-q_2e_1+q_2b)+q_0c-b,\\
\frac{d q_0}{dt}=-q_2+\frac{1}{2}e_3\big(q_2^2-1\big) +c\left(q_1q_2+\frac{1}{2}e_1\big(1-q_2^2\big)\right),\\
\frac{d q_1}{dt}=-q_1q_2b+\frac{1}{2}\big(q_2^2-1\big)(e_2+be_1+c),\\
\frac{d q_2}{dt}=\big(q_2^2-1\big)\left(e_1-\frac{1}{2}b\right)-q_1q_2.\label{rumsys2}
\end{gather}
We observe that the equations (\ref{rumsys1})--(\ref{rumsys2}) f\/ix only six of the nine free parameters introduced to def\/ine the matrices $L$ and $B$ in (\ref{lax2}) and (\ref{lax3}) respectively. The parameters that still need to be f\/ixed are $b$, $c$ and $d$, or equivalently, $b$, $c$ and $U$ introduced in~(\ref{Udef}).

The system of equations (\ref{rumsys1})--(\ref{rumsys2}) has a set of integrals of motions that was obtained in~\cite{Rumanov}. In order to def\/ine these integrals let us introduce the auxiliary functions
\begin{gather}\label{r2}
r_2=\dfrac{q_2^2-1}{4}\big(e_1^2-e_2\big)-\frac{1}{2}e_1q_1q_2+\frac{1}{2}q_2q_0+\frac{1}{4}q_1^2,
\\
\label{r1}
r_1=\dfrac{q_2^2-1}{4}(e_3-e_2e_1)+\frac{1}{2}e_2q_1q_2-\frac{1}{2}q_1q_0,
\end{gather}
and
\begin{gather}\label{r0}
r_0=\frac{q_0^2}{4}+e_1e_3\frac{q_2^2-1}{4}-\frac{1}{2}e_3q_1q_2.
\end{gather}
Then the quantities
\begin{gather*}
I_0 = 2r_0 + U -e_1q_2+2q_1, \qquad
I_1 = 2r_1 -1 -q_2,
\end{gather*}
and
\begin{gather*}
I_2 = 2r_2 +t
\end{gather*}
are the integrals of the system (\ref{rumsys1})--(\ref{rumsys2})~\cite{Rumanov}.
A key observation now is the following statement.
\begin{Proposition}[\cite{Rumanov}] The Lax pair \eqref{lax1} implies the differential identity
\begin{gather*}
 3(\Psi_{11})_t+(\Psi_{11})_{xx}+\big(t-x^2\big)(\Psi_{11})_x\\
\qquad {} = x^2\left(r_2+\frac{t}{2}\right)\Psi_{11} +x(3b-2e_1)\Psi_{21}+x \left(r_1-\frac{1}{2}-\frac{q_2}{2}\right)\Psi_{11}\\
\qquad \quad {} +(e_2+3c)\Psi_{21}+\left(\frac{U}{2} + \frac{3}{2}q_2(b-e_1) + q_1 + r_0\right)\Psi_{11}.
\end{gather*}
\end{Proposition}
An immediate consequence of this proposition is that the function
\begin{gather*}
F(x,t;\beta=6) = \Psi_{11}\bigl(3^{2/3}t,3^{1/3}x\bigr)
\end{gather*}
satisfy the Bloemendal--Vir\'ag equation~(\ref{bv1}) for $\beta =6$ if the following constraints are imposed on the functional parameters of the Lax pair~(\ref{lax1})
\begin{gather}
r_2=-\frac{t}{2},\label{r2eq}\\
r_1=\frac{1}{2}+\frac{q_2}{2},\label{r1eq}\\
b=\frac{2}{3}e_1,\label{Btileq}\\
c=-\frac{1}{3}e_2,\label{btileq}\\
\frac{U}{2} + \frac{3}{2}q_2(b-e_1) + q_1 + r_0=0.\label{UtilB}
\end{gather}
Constraints (\ref{r2eq}) and (\ref{r1eq}) are the restrictions on the integrals $I_1$ and $I_2$:
\begin{gather}\label{restint}
I_1 = I_2 = 0.
\end{gather}
Constraint (\ref{UtilB}) can be also easily achieved; indeed, this is just a formula for the yet free functional parameter~$U$. Equations~(\ref{Btileq}) and~(\ref{btileq}) are the genuine extra conditions on the functional parameters of the Lax pair (\ref{lax1}) which do not follow from the zero curvature equation~(\ref{lax22}). They make the relation of Rumanov's Lax pair~(\ref{lax1}) to the Bloemendal--Vir\'ag equation~(\ref{bv1}) with $\beta=6$ not as straightforward as the relation of the standard Painlev\'e~II Lax pair to the Bloemendal--Vir\'ag equation~(\ref{bv1}) with $\beta=2$. We will come back to this issue in Section~\ref{gauge}.

We conclude this section by noticing that together with (\ref{Btileq}), equation (\ref{UtilB}) implies that the third remaining Rumanov's integral is also zero,
\begin{gather*}
I_0 = 0.
\end{gather*}

\section{WKB analysis of Rumanov's Lax pair}

In this section we present the large $x$ asymptotic analysis of the solution $\Psi(x,t)$ of the Lax pair~(\ref{lax1}). Our consideration will be formal. The goal of this section is twofold. First, we want to explain the WKB-meaning of Rumanov's integrals $I_1$ and $I_2$. Secondly, the formulae obtained here will serve as a motivation for the principal constructions of Section~\ref{gauge} which in turn will allow us in Section~\ref{sect6} to obtain our main result~-- the proof of Conjecture~\ref{rumcon}.

The formal large $x$ asymptotics of the function $\Psi(x)$ is given by the following classical \cite{W} WKB-ansatz
\begin{gather}\label{wkbansatz}
\Psi_{\rm WKB}(x)=T(x)\exp\left[\int \Lambda dx-\int\operatorname{diag} \left(T^{-1}\frac{dT}{dx}\right)dx\right],
\end{gather}
where the diagonal matrix $\Lambda$ and the invertible matrix $T$ are taken from the spectral decomposition of the matrix $L$,
\begin{gather*}
L=T\Lambda T^{-1}, \qquad \Lambda = \begin{pmatrix}\lambda_+&0\\ 0&\lambda_-\end{pmatrix}.
\end{gather*}
The eigenvalues $\lambda_{\pm}$ are the roots of the characteristic equation
\begin{gather*}
\det(\lambda-L)\equiv \left(\lambda-\frac{x^2-t}{2}\right)^2-\left(\frac{x^4}{4}+r_2x^2+r_1x+r_0\right)=0,
\end{gather*}
where $r_2$, $r_1$ and $r_0$ are exactly the same as in (\ref{r2})--(\ref{r0}). This means that the eigenvalues are
\begin{gather}\label{eigen}
\lambda_{\pm}=\frac{x^2-t}{2}\pm \mu,\qquad \mu=\sqrt{\frac{x^4}{4}+r_2x^2+r_1x+r_0}=\frac{x^2}{2}+r_2+\frac{r_1}{x}++\frac{r_0-r_2^2}{x^2}+\cdots,\!\!\!
\end{gather}
and we also have
\begin{gather*}
T=\begin{pmatrix}Q(x)+\mu&Q(x)-\mu\\P(x)&P(x)\end{pmatrix}\\
\hphantom{T}{} =\frac{1}{2}\begin{pmatrix}x^2(1+q_2)-q_1x+ q_0+2r_2 + \cdots &(q_2-1)x^2-q_1x+q_0-2r_2 + \cdots\vspace{1mm}\\-x\dfrac{q_2^2-1}{2}+q_2q_1-e_1\dfrac{q_2^2-1}{2}&-x\dfrac{q_2^2-1}{2}+q_2q_1-e_1\dfrac{q_2^2-1}{2}\end{pmatrix},
\end{gather*}
where
\begin{gather*}
P(x)=-\frac{(q_2^2-1)(x+e_1)-2q_2q_1}{4},\qquad Q(x)=\frac{q_2x^2-q_1x+q_0}{2}.
\end{gather*}
In particular, we have that
\begin{gather}\label{Texp}
T(x)=\begin{pmatrix}x^2&0\\0&x\end{pmatrix}\begin{pmatrix}\dfrac{1+q_2}{2}&\dfrac{q_2-1}{2}\vspace{1mm}\\ \dfrac{1-q_2^2}{4}&\dfrac{1-q_2^2}{4}\end{pmatrix}\left[I+ \frac{\widehat{M}}{x} +\cdots\right],
\end{gather}
where
\begin{gather*}
\widehat{M} = \begin{pmatrix}
\tilde{\alpha} & \tilde{\alpha} \\ \alpha & \alpha
\end{pmatrix}, \qquad \tilde{\alpha}=\frac{1-q_2}{2(1+q_2)}(-q_1+(1+q_2)e_1),\qquad \alpha=\frac{1+q_2}{2(1-q_2)}(q_1+(1-q_2)e_1).
\end{gather*}
Plugging the estimates (\ref{eigen}) and (\ref{Texp}) into the right hand side of equation (\ref{wkbansatz}), we arrive at the following expansion as $x \to \infty$
\begin{gather}\label{wkb1}
\Psi_{\rm WKB}(x)=x^{\frac{\sigma_3}{2}}\begin{pmatrix}\dfrac{1+q_2}{2}&\dfrac{q_2-1}{2}\vspace{1mm}\\ \dfrac{1-q_2^2}{4}&\dfrac{1-q_2^2}{4}\end{pmatrix}\left[I+\frac{M}{x}+\cdots\right]e^{\frac{x^3}{6}-\frac{tx}{2}}
e^{\left(\frac{x^3}{6}+r_2x\right)\sigma_3+\nu\log x\sigma_3},
\end{gather}
where
\begin{gather*}
\nu=r_1-\frac{q_2}{2},\qquad
M=\begin{pmatrix}
r_2^2-r_0 - \dfrac{q_1}{2}&\tilde{\alpha}\\
\alpha&r_0-r_2^2 + \dfrac{q_1}{2}\end{pmatrix}.
\end{gather*}
Note that
\begin{gather*}
r_2 = -\frac{t}{2} + \frac{1}{2}I_2 \qquad \text{and}\qquad \nu = \frac{1}{2} + \frac{1}{2}I_1.
\end{gather*}
This shows the role of the integrals $I_2$ and $I_1$: Integral $I_2$ determines the exponential function
describing the essential singularity of the function $\Psi(x)$ at $x =\infty$, while the
integral $I_1$ determines the formal exponent $\nu$ at $x =\infty$. Also, taking into account
conditions (\ref{restint}) we conclude that
\begin{gather*}
r_2 = -\frac{t}{2} \qquad\text{and}\qquad \nu = \frac{1}{2},
\end{gather*}
and rewrite expansion (\ref{wkb1}) as
\begin{gather}\label{wkb11}
\Psi_{\rm WKB}(x)=x^{\frac{\sigma_3}{2}}\begin{pmatrix}\dfrac{1+q_2}{2}&\dfrac{q_2-1}{2}\vspace{1mm}\\ \dfrac{1-q_2^2}{4}&\dfrac{1-q_2^2}{4}\end{pmatrix}\left[I+\frac{M}{x}+\cdots\right]e^{\frac{x^3}{6}-\frac{tx}{2}}
e^{\left(\frac{x^3}{6}-\frac{tx}{2}\right)\sigma_3+\frac{1}{2}\log x\sigma_3}.
\end{gather}
Observe f\/inally that expansion (\ref{wkb11}) with the proper modif\/ication of the matrix coef\/f\/icient~$M$, can be written in the form
\begin{gather}\label{wkb2}
\Psi_{\rm WKB}(x)=\begin{pmatrix}
\dfrac{1}{2}(1+q_2)x-\alpha&-1\\
\dfrac{1}{4}(1-q_2^2)&0
\end{pmatrix}
\left[I+\frac{M_0}{x}+\cdots\right]e^{\frac{x^3}{6}-\frac{tx}{2}}
e^{\left(\frac{x^3}{6}-\frac{tx}{2}\right)\sigma_3},
\end{gather}
where
\begin{gather*}
M_0=\begin{pmatrix}
r_2^2-r_0-\dfrac{q_1}{2}+\alpha&1\vspace{1mm}\\
\alpha\left(r_0-r_2^2+\dfrac{q_1}{2}-\alpha\right)&r_0-r_2^2+\dfrac{q_1}{2}-\alpha
\end{pmatrix}.
\end{gather*}
The formal series
\begin{gather*}
\left[I+\frac{M_0}{x}+\cdots\right]e^{\left(\frac{x^3}{6}-\frac{tx}{2}\right)\sigma_3},
\end{gather*}
characterizes the essential singularity of the canonical solutions of the auxiliary linear system corresponding to the second Painlev\'e equation (see, e.g., \cite{FIKN}). Therefore, formula (\ref{wkb2}) suggests that Rumanov's Lax pair should be gauge equivalent to the standard Painlev\'e II Lax pair. The exact description of this gauge equivalence will be given in the next section.

\section{The gauge transformation to the Painlev\'e~II Lax pair}\label{gauge}

Let $\Psi_0(x, t)$ be a (fundamental) solution of the Flaschka--Newell Painlev\'e II Lax pair, i.e.,
\begin{gather}\label{back1}
\frac{d\Psi_0}{dx}=\hat{L}_0\Psi_0,\qquad \frac{d\Psi_0}{dt}=\hat{B}_0\Psi_0,
\end{gather}
where the matrices $\hat{L}_0$ and $\hat{B}_0$ are
\begin{gather*}
\hat{L}_0=\frac{x^2}{2}\sigma_3+x\begin{pmatrix}0&u\\ u&0\end{pmatrix}+\begin{pmatrix}\delta&w\\-w&-\delta\end{pmatrix}
\end{gather*}
and
\begin{gather*}
\hat{B}_0=-\frac{x}{2}\sigma_3-\begin{pmatrix}0&u\\ u&0\end{pmatrix}.
\end{gather*}
The parameters $u$, $w$, $\delta$ are related by the equations
\begin{gather*}
w=-u_t, \qquad \delta = -\frac{t}{2} -u^2.
\end{gather*}

The compatibility condition of the pair (\ref{back1}) is the second Painlev\'e equation (cf.~(\ref{p2int}))
\begin{gather}\label{P2}
u_{tt} = tu +2u^3.
\end{gather}
We shall discuss the particular choice of the Painlev\'e II function $u(t)$ and of the $\Psi_0$ -- function later in Section~\ref{sect6}.

Taking a hint from (\ref{wkb2}), we put
\begin{gather*}
\Psi(x,t) := e^{\frac{x^3}{6} -\frac{xt}{2}}\kappa(t)R(x,t)\psi^{\sigma_3}(t)\Psi_0(x,t),
\end{gather*}
where
\begin{gather*}
R(x,t) = \begin{pmatrix}p(t)x -\alpha(t)&-1\\q(t)&0\end{pmatrix}.
\end{gather*}
Here, at the moment, $\kappa$, $\psi$, $p$, $q$ and $\alpha$ are free functional parameters. The proof of the following statement is straightforward though a bit tedious.
\begin{Proposition} The function $\Psi(x,t)$ satisfies the Lax pair
\begin{gather*}
\frac{d\Psi}{dx}=L\Psi,\qquad \frac{d\Psi}{dt}=B\Psi,
\end{gather*}
where the matrices $L$ and $B$ are
\begin{gather}\label{Lback}
L = R\psi^{\sigma_3}\hat{L}_0\psi^{-\sigma_3}R^{-1} + R_xR^{-1} + \left(\frac{x^2}{2} - \frac{t}{2}\right)I \equiv x^3J +x^2{ L}_2 + x{ L}_1 + { L}_0,
\end{gather}
and
\begin{gather}\label{Bback}
B= R\psi^{\sigma_3}\hat{B}_0\psi^{-\sigma_3}R^{-1} + R_tR^{-1} - \frac{x}{2}I +\frac{\kappa_t}{\kappa}I +\frac{\psi_t}{\psi} R\sigma_3R^{-1}
 \equiv -x^2J +x{B}_1 + {B}_0,
\end{gather}
and the matrix coefficients ${L}_{k}$ and ${B}_k$ are given in terms of the Painlev\'e function $u(t)$ and $($still free$)$ functional parameters $\kappa$, $\psi$, $p$, $q$ and $\alpha$ by the following equations:
\begin{gather}\label{Jback}
J = \begin{pmatrix}0&\dfrac{p}{q}(1 + pu\psi^2)\\0&0\end{pmatrix},
\\
\label{L2back}
{L}_2 = \begin{pmatrix}-pu\psi^2&-\dfrac{\alpha}{q} -\dfrac{2\alpha p}{q}u\psi^{2} + \dfrac{p^2}{q}w\psi^2\vspace{1mm}\\
0&1+ pu\psi^2 \end{pmatrix},
\\
\label{L1back}
{L}_1 = \begin{pmatrix} (u\alpha -pw)\psi^2&\dfrac{\alpha^2}{q}u\psi^2 -\dfrac{1}{q}v\psi^{-2} + \dfrac{2p}{q}\delta
-\dfrac{2\alpha p}{q}w\psi^2\vspace{1mm}\\
-qu\psi^2&-(u\alpha -pw)\psi^2\end{pmatrix},
\\
\label{L0back}
{L}_0 = \begin{pmatrix}w\alpha \psi^2 -\delta -\dfrac{t}{2}&-\dfrac{2\alpha\delta}{q}+\dfrac{\alpha^2}{q}w\psi^{2} - \dfrac{1}{q}y\psi^{-2}
+\dfrac{p}{q}\vspace{1mm}\\
-qw\psi^2&-w\alpha \psi^2 +\delta-\frac{t}{2}\end{pmatrix},
\end{gather}
and
\begin{gather}\label{B1back}
{B}_1 = \begin{pmatrix}pu\psi^2&\dfrac{\alpha}{q} +\dfrac{2\alpha p}{q}u\psi^{2} + \dfrac{p_t}{q} +2\dfrac{\psi_t}{\psi}\dfrac{p}{q}\vspace{1mm}\\
0&-1- pu\psi^2 \end{pmatrix},
\\
\label{B0back}
{B}_0 = \begin{pmatrix}\dfrac{\kappa_t}{\kappa} - \dfrac{\psi_t}{\psi} -\alpha u\psi^2
 &-\dfrac{\alpha^2}{q}u\psi^2 +\dfrac{1}{q}v\psi^{-2} -\dfrac{\alpha_t}{q}-2\dfrac{\psi_t}{\psi}\dfrac{\alpha}{q}\vspace{1mm}\\
qu\psi^2&\dfrac{\kappa_t}{\kappa} + \dfrac{\psi_t}{\psi} +\alpha u\psi^2 +\dfrac{q_t}{q}\end{pmatrix}.
\end{gather}
\end{Proposition}

We want now to match the formulae (\ref{Lback})--(\ref{B0back}) with the formulae (\ref{lax2}), (\ref{lax3}). From equa\-tion~(\ref{Jback}) we arrive at the f\/irst restriction
\begin{gather}\label{restback1}
\frac{p}{q}\big(1 + pu\psi^2\big) = 1.
\end{gather}
Also, taking again the hint from (\ref{wkb2}) we choose $p$ and $q$ in the form
\begin{gather*}
p= \frac{1+q_2}{2}, \qquad q=\frac{1-q^2_2}{4},
\end{gather*}
where $q_2$ is a new free functional parameter. With this choice, equation (\ref{restback1}) transforms into the relation
\begin{gather*}
\frac{1+q_2}{2}u\psi^2 = - \frac{1+q_2}{2},
\end{gather*}
and hence the formula for $\psi$:
\begin{gather*}
\psi^{2} = -\frac{1}{u}.
\end{gather*}
Therefore, the number of free parameters is reduced from f\/ive to three: $q_2$, $\alpha$, $\kappa$, and the f\/inal formulae for the matrix coef\/f\/icients ${L}_{k}$ and ${B}_k$ are the following:
\begin{gather*}
J = \begin{pmatrix}0&1\\0&0\end{pmatrix},
\\ 
{L}_2 = \begin{pmatrix}\dfrac{1+q_2}{2}&\dfrac{4\alpha q_2}{1-q^2_2} + \dfrac{u_t}{u}\dfrac{1+q_2}{1-q_2}\vspace{1mm}\\
0&\dfrac{1-q_2}{2} \end{pmatrix},
\\ 
{L}_1 = \begin{pmatrix} -\alpha -\dfrac{u_t}{u}\dfrac{1+q_2}{2}&
\dfrac{4}{1-q^2_2}\left(-\alpha^2 +u^2 + (1+q_2)\delta -(1+q_2)\alpha\dfrac{u_t}{u}\right)\vspace{1mm}\\
\dfrac{1-q_2^2}{4}&\alpha +\dfrac{u_t}{u}\dfrac{1+q_2}{2}\end{pmatrix},
\\ 
{L}_0 = \begin{pmatrix}\alpha \dfrac{u_t}{u} -\delta-\dfrac{t}{2}&
\dfrac{4}{1-q^2_2}\left(-2\alpha\delta+\alpha^2\dfrac{u_t}{u} + u_tu +\dfrac{1+q_2}{2}\right)\vspace{1mm}\\
-\dfrac{1-q_2^2}{4}\dfrac{u_t}{u}&-\alpha \dfrac{u_t}{u} +\delta-\dfrac{t}{2}\end{pmatrix},
\end{gather*}
and
\begin{gather*}
{B}_1 = \begin{pmatrix}-\dfrac{1+q_2}{2}&\dfrac{4}{1-q^2_2}\left(
-\alpha q_2 + \dfrac{q_{2t}}{2} -\dfrac{u_t}{u}\dfrac{1+q_2}{2}\right)\vspace{1mm}\\
0&-\dfrac{1-q_2}{2} \end{pmatrix},
\\ 
{B}_0 = \begin{pmatrix}\dfrac{\kappa_t}{\kappa} +\dfrac{1}{2} \dfrac{u_t}{u} +\alpha
 &\dfrac{4}{1-q^2_2}\left(\alpha^2 - u^2 -\alpha_t+\alpha \dfrac{u_t}{u}\right)\vspace{1mm}\\
-\dfrac{1-q^2_2}{4}&\dfrac{\kappa_t}{\kappa} -\dfrac{1}{2} \dfrac{u_t}{u} -\alpha -\dfrac{2q_2q_{2t}}{1-q^2_2}\end{pmatrix}.
\end{gather*}
Comparing these formulae with Rumanov's Lax pair~(\ref{lax2}),~(\ref{lax3}) we see that our~$L$ and~$B$ have exactly the same structure with Rumanov's parameters, $q_1$, $q_0$, $e_1$, $e_2$, $e_3$, $b$, $c$, ${a}$, and ${d}$ given in terms of our parameters $\alpha$, $q_2$, and $\kappa$ by the equations
\begin{gather}\label{q1back}\allowdisplaybreaks
q_1 = 2\alpha +\frac{u_t}{u}(1+q_2),
\\ \label{q0back}
q_0 = 2\alpha \frac{u_t}{u} -2\delta,
\\ \label{e1back}
e_1 = -\frac{4\alpha q_2}{1-q^2_2} -\frac{u_t}{u}\frac{1+q_2}{1-q_2},
\\ \label{e2back}
e_2 = \frac{4}{1-q^2_2}\left(-\alpha^2 +u^2 + (1+q_2)\delta -(1+q_2)\alpha\frac{u_t}{u}\right),
\\ \label{e3back}
e_3 = - \frac{4}{1-q^2_2}\left(-2\alpha\delta+\alpha^2\frac{u_t}{u} + u_tu +\frac{1+q_2}{2}\right),
\\\label{tilaback}
{a} = \frac{\kappa_t}{\kappa} +\frac{1}{2} \frac{u_t}{u} +\alpha,
\\ \label{tildback}
{d} = \frac{\kappa_t}{\kappa} -\frac{1}{2} \frac{u_t}{u} -\alpha -\frac{2q_2q_{2t}}{1-q^2_2},
\\ \label{tilBback}
b = \frac{4}{1-q^2_2}\left(
-\alpha q_2 + \frac{q_{2t}}{2} -\frac{u_t}{u}\frac{1+q_2}{2}\right),
\\ \label{tilbback}
c = \frac{4}{1-q^2_2}\left(\alpha^2 - u^2 -\alpha_t+\alpha \frac{u_t}{u}\right).
\end{gather}

It is worth noticing that the relation
\begin{gather*}
(L_0){_{21}} \equiv -\frac{1-q_2^2}{4}\frac{u_t}{u} = e_1\frac{1-q_2^2}{4} + \frac{q_1q_2}{2},
\end{gather*}
which is present in (\ref{lax2}) follows from (\ref{q1back}), (\ref{e1back}) automatically. Also, automatically, we have that
\begin{gather*}
\alpha=\dfrac{1+q_2}{2(1-q_2)}(q_1+e_1(1-q_2)),
\end{gather*}
and relation (\ref{atild}). Moreover, the following statement is true (the proof is again straightforward).
\begin{Proposition} Let $r_2$ and $r_1$ are defined according to \eqref{r2} and \eqref{r1} where all the parameters $e_1$, $e_2$, $e_3$, $q_1$, and $q_0$ are given as functions of $q_2$, $\alpha$, and $u$ according to \eqref{q1back}--\eqref{e3back}. Then,
\begin{gather*}
r_2 = -\frac{t}{2} \qquad\text{and}\qquad r_1 = \frac{q_2}{2} +\frac{1}{2}\qquad \forall\, q_2,\, \alpha,\, u.
\end{gather*}
\end{Proposition}
From this Proposition it follows that the constraints (\ref{r2eq}) and (\ref{r1eq}) are satisf\/ied automatically while the equations (\ref{Btileq}), (\ref{btileq}), and (\ref{UtilB}) must be imposed if we want the function
\begin{gather*}
F(x,t;\beta=6) : = \Psi_{11}\bigl(3^{1/3}x, 3^{2/3}t\bigr)
\end{gather*}
to satisfy the Bloemendal--Vir\'ag equation. Equations (\ref{Btileq}) and (\ref{btileq}) yield the ODEs for $q_2$ and $\alpha$ while equation~(\ref{UtilB}) produces formula for~$\kappa$. In the next sections we will analyze (\ref{Btileq}), (\ref{btileq}), and (\ref{UtilB}).

\section[ODEs for $q_2$ and $\alpha$]{ODEs for $\boldsymbol{q_2}$ and $\boldsymbol{\alpha}$} \label{S5}
Consider f\/irst equation (\ref{Btileq}), i.e.,
\begin{gather*}
b = \frac{2}{3}e_1.
\end{gather*}
Substituting here (\ref{tilBback}) and (\ref{e1back}) we arrive at the following dif\/ferential equation for $q_2(t)$
\begin{gather}\label{q2dif}
q_{2t} = q_2\left(\frac{2}{3}\alpha + \frac{u_t}{u}\frac{2-q_2}{3}\right) +\frac{u_t}{u}\frac{2-q_2}{3}.
\end{gather}

Next, we look at equation (\ref{btileq}), i.e.,
\begin{gather*}
c = -\frac{1}{3}e_2.
\end{gather*}
Substituting here (\ref{tilbback}) and (\ref{e2back}) we arrive at the following dif\/ferential equation for $\alpha(t)$:
\begin{gather}\label{alpha2dif}
\alpha_{t} = \alpha\left(\frac{2}{3}\alpha + \frac{u_t}{u}\frac{2-q_2}{3}\right) - \frac{t}{6}(1+q_2)- \frac{u^2}{3}(3+q_2).
\end{gather}

Note that (\ref{q2dif}) can be rewritten as
\begin{gather}\label{q2eq2}
q_{2t} = \frac{2}{3}\alpha q_2 + \frac{u_t}{u}\frac{(1+q_2)(2-q_2)}{3},
\end{gather}
which, in particular, yields the following formula for $\alpha$ in terms of $q_2$
\begin{gather}\label{alphaq2}
\alpha = \frac{3}{2}\frac{q_{2t}}{q_2} - \frac{u_t}{u}\frac{(1+q_2)(2-q_2)}{2q_2},
\end{gather}
and, in turn, allows us to transform the system (\ref{q2dif}), (\ref{alpha2dif}) of two f\/irst order ODEs to a single second order ODE for the function $q_2$:
\begin{gather}
q_{2tt} = \frac{2q_{2t}}{q_{2}}\left(q_{2t} - \frac{u_t}{u}\right) +\frac{4}{9}\left(q^2_2 -\frac{3}{2}\right)
\left(\frac{u^2_t}{u^2} -t -2u^2\right)
 -\frac{2}{9}q_2\left(3\frac{u^2_t}{u^2} -t\right) + \frac{4}{9q_2}\frac{u^2_t}{u^2}.\label{q2eq3}
\end{gather}
Put
\begin{gather*}
\eta = \frac{2\alpha}{q_2 - 1} -\frac{u^2}{\omega} -\frac{u_t}{u}\frac{1+q_2}{1-q_2}, \qquad \omega = u^4 + tu^2 - u^2_t.
\end{gather*}
Then equation (\ref{q2eq3}) transforms to the following equation for the function~$\eta$
\begin{gather}\label{etaeq}
9\eta_{tt} + 9\eta\eta_t +\eta^3 + P(t)\eta + Q(t) = 0,
\end{gather}
where
\begin{gather*}
P(t) = 12\left(\frac{u^2}{\omega} -\omega \right)_t - 4t, \qquad Q(t) = \frac{2}{3}P_t(t) + \frac{2}{3}.
\end{gather*}
This is equation~(1.14) of Rumanov's paper.

Using (\ref{q1back}), we can pass from the pair $(q_2, \alpha)$ to the pair $(q_2, q_1)$. For the new unknowns, equations (\ref{q2dif}), (\ref{alpha2dif}) transform to the following pair of equations
\begin{gather}\label{q2new}
q_{2t} = \frac{1}{3}q_1q_2 +\frac{2}{3}\frac{u_t}{u}\big(1-q^2_2\big)
\end{gather}
and
\begin{gather}\label{q1new}
q_{1t} = -\frac{2}{3}\frac{u_t}{u}q_1q_2 +\frac{1}{3}q^2_1
+\frac{2}{3}\left(t - \frac{u^2_t}{u^2} \right) + \frac{2}{3}\left(\frac{u_t}{u}\right)_tq_2.
\end{gather}
These are equations (3.17) and (3.18) of paper~\cite{Rumanov} (we note that our $u(t)$ is Rumanov's~$q(t)$). A very interesting fact shown in~\cite{Rumanov} is that these equations can be linearized. Consider the following $3\times 3$ linear system for the new functions $\mu_{\pm}$ and $\nu$,
\begin{gather}\label{mu+}
\mu_{+t} = \frac{2}{3}\frac{u_t}{u}\mu_+ -\frac{1}{3}\nu,
\\ \label{mu-}
\mu_{-t} = -\frac{2}{3}\frac{u_t}{u}\mu_- +\frac{1}{3}\nu,
\\ \label{nu}
\nu_{t} = \frac{2}{3}u^2\mu_- +\frac{2}{3}\frac{\omega}{u^2}\mu_+
\end{gather}
Then, as it is shown in Lemma~2 of \cite{Rumanov}, the formulae
\begin{gather}\label{q2mu}
q_2 = \frac{\mu_+ +\mu_-}{\mu_+ -\mu_-} \qquad\text{and}\qquad q_1 = \frac{2\nu}{\mu_+ -\mu_-}
\end{gather}
determine solution of the nonlinear system (\ref{q2new}), (\ref{q1new}). It is worth reproducing the proof of Rumanov's lemma.

Put
\begin{gather*}
\mu:= \frac{\mu_+ + \mu_-}{2} \qquad\text{and}\qquad \chi:= \frac{\mu_+ - \mu_-}{2}.
\end{gather*}
Then, the system (\ref{mu+})--(\ref{nu}) can be rewritten as
\begin{gather}\label{mu}
\mu_{t} = \frac{2}{3}\frac{u_t}{u}\chi,
\\ \label{chi}
\chi_{t} = \frac{2}{3}\frac{u_t}{u}\mu -\frac{1}{3}\nu,
\\ \label{nu2}
\nu_{t} = \frac{2}{3}\left(\frac{u_t}{u}\right)_t\mu +\frac{2}{3}\left(t - \frac{u^2_t}{u^2} \right)\chi,
\end{gather}
where we have taken into account the relation (consequence of the Painlev\'e equation)
\begin{gather*}
u^2 + \frac{\omega}{u^2} = \left(\frac{u_t}{u}\right)_t.
\end{gather*}
Simultaneously, formulae (\ref{q2mu}) become the formulae
\begin{gather}\label{q2muchi}
q_2 = \frac{\mu}{\chi} \qquad\text{and}\qquad q_1 = \frac{\nu}{\chi},
\end{gather}
assuming that $\mu$, $\chi$, and $\nu$ satisfy (\ref{mu})--(\ref{nu2}). Then, from (\ref{q2muchi}) we would have that
\begin{gather*}
q_{2t} = \frac{\mu_t}{\chi} - \frac{\mu}{\chi^2}\chi_t =
\frac{2}{3}\frac{u_t}{u} - \frac{\mu}{\chi}
\left(\frac{2}{3}\frac{u_t}{u}\frac{\mu}{\chi} - \frac{1}{3}\frac{\nu}{\chi}\right)
=\frac{2}{3}\frac{u_t}{u} - q_2\left(\frac{2}{3}\frac{u_t}{u}q_2 - \frac{1}{3}q_1\right),
\end{gather*}
which is identical to~(\ref{q2new}). Similarly, we have that
\begin{gather*}
q_{1t} = \frac{\nu_t}{\chi} - \frac{\nu}{\chi^2}\chi_t =
\frac{2}{3}\left(\frac{u_t}{u}\right)_t\frac{\nu}{\chi} +\frac{2}{3}\left(t - \frac{u^2_t}{u^2} \right)-\frac{\nu}{\chi}
\left(\frac{2}{3}\frac{u_t}{u}\frac{\mu}{\chi} - \frac{1}{3}\frac{\nu}{\chi}\right)\\
\hphantom{q_{1t}}{}
=\frac{2}{3}\left(\frac{u_t}{u}\right)_tq_1 +\frac{2}{3}\left(t - \frac{u^2_t}{u^2} \right)-q_1
\left(\frac{2}{3}\frac{u_t}{u}q_2 - \frac{1}{3}q_1\right),
\end{gather*}
which is (\ref{q1new}).

The possibility to linearize the auxiliary nonlinear system (\ref{q2new}),~(\ref{q1new}), and hence the auxiliary Lax pair constraints~(\ref{Btileq}) and (\ref{btileq}), is, of course, a very important observation. However, this, by itself, does not put the $\beta=6$ case on the same footing as the case $\beta =2$. Indeed, the original Bloemendal--Vir\'ag equation (\ref{bv1}) is already a linear dif\/ferential equation. True, it is a~PDE equation and not a system of ODE equations. Still, to make Rumanov's approach a~really ef\/f\/icient scheme, one needs to put the system (\ref{mu+})--(\ref{nu}) in an integrable context. In fact, this could be crucial for proving Conjectures~\ref{rumcon2} and~\ref{bemncon1}. We shall say more about this issue in the concluding section of the paper.

One of the possible ways to ``integrate'' equations (\ref{mu+})--(\ref{nu}) would be to express the functions $\mu_{\pm}(t)$ and $\nu(t)$ in terms of the solution $\Psi_{0}(x,t)$ of the Painlev\'e II Lax pair~(\ref{back1}). In fact, it is not necessary to try to reduce (\ref{mu+})--(\ref{nu}) to (\ref{back1}).
Any alternative Lax pairs for the second Painlev\'e equation would do. We shall analyze the possibility to connect the system (\ref{mu+})--(\ref{nu}) to one of the known Lax pairs for Painlev\'e II in our future publication.

\begin{Remark} As it was pointed out to us by A.~Prokhorov, equation (\ref{etaeq}) can be linearized even more directly. One can easy check that the substitution
\begin{gather*}
\eta = 3\frac{d\ln f}{dt},
\end{gather*}
transforms (\ref{etaeq}) to the following third order linear equation for the function $f(t)$
\begin{gather*}
27f_{ttt} + 3P(t)f_t + Q(t)f = 0.
\end{gather*}
This fact by itself does not lead, however, to the proof of Conjecture~\ref{bemncon1}~-- see discussion in the concluding section of the paper.
\end{Remark}

\section[Formulae for $r_0$, $\kappa$, and $U(t)$. The third Rumanov's integral]{Formulae for $\boldsymbol{r_0}$, $\boldsymbol{\kappa}$, and $\boldsymbol{U(t)}$. The third Rumanov's integral}

In this short section we present expressions for $\kappa$ and $U$ which follow from (\ref{UtilB}) and also establish that, as expected, the third Rumanov's integral, $I_0$ vanishes.

Substituting into (\ref{r0}) equations (\ref{q1back}), (\ref{q0back}), (\ref{e1back}), and (\ref{e3back}), we arrive at the following formula for~$r_0$
\begin{gather}\label{r011}
r_0 = \omega + \frac{t^2}{4} -\frac{u_t}{u}\frac{1+q_2}{2},
\end{gather}
where
\begin{gather*}
\omega:= u^4 + tu^2 - u_t^{2}.
\end{gather*}
Note that $-\frac{1}{2}\omega$ is the Hamiltonian of the second Painlev\'e equation (\ref{P2}), and also that
\begin{gather*}
\omega_t = u^2.
\end{gather*}
In its turn, equation (\ref{r011}) together with the equations (\ref{tilaback}) and (\ref{q1back}), after having been substituted into (\ref{UtilB}), yields the formula for~$\kappa$
\begin{gather}\label{kappa11}
\frac{\kappa_t}{\kappa} = -\frac{1}{3}\omega -\frac{2}{3}\alpha -\frac{u_t}{u}\frac{1-2q_2}{6}.
\end{gather}
At the same time, from (\ref{tilaback}), (\ref{tildback}), (\ref{atild}) and (\ref{Udef}), we have that
\begin{gather*}
U(t) = 6\frac{\kappa_t}{\kappa} -6\frac{q_2q_{2t}}{1-q^2_2} - \frac{t^2}{2},
\end{gather*}
or, taking into account (\ref{kappa11}),
\begin{gather*}
U(t) = -2\omega -4\alpha -\frac{u_t}{u}(1-2q_2) -6\frac{q_2q_{2t}}{1-q^2_2} - \frac{t^2}{2}.
\end{gather*}
Furthermore, taking into account ODE (\ref{q2eq2}), we can transform the last equation into the equation
\begin{gather}\label{U4}
U(t) = -2\omega -\frac{4}{1-q^2_2} \alpha -\frac{u_t}{u}\frac{1+q_2}{1-q_2} - \frac{t^2}{2}.
\end{gather}

Now, we can check the third Rumanov's integral, i.e.,
\begin{gather*}
I_0 = \frac{q^2_0}{2} +e_1e_2\frac{q^2_2-1}{2} -e_3q_1q_2 + U -e_1q_2 +2q_1.
\end{gather*}
This must be zero. We notice that it is related to our function $r_0$ by the equation
\begin{gather*}
I_0(t) = 2r_0 + U -e_1q_2 + 2q_1.
\end{gather*}
Substituting here (\ref{r011}), (\ref{U4}) and (\ref{q1back}) we check that indeed
\begin{gather*}
I_0 (t) \equiv 0.
\end{gather*}

\section[$\beta=6$ Tracy--Widom function. The proof of Conjecture \ref{rumcon}]{$\boldsymbol{\beta=6}$ Tracy--Widom function. The proof of Conjecture \ref{rumcon}}\label{sect6}

The results of the previous sections can be formulated as the following theorem
\begin{Theorem} Let $u(t)$ and $\Psi_0(x,t)$ be the solutions of the second Painlev\'e equation~\eqref{P2}, and of the Lax pair \eqref{back1}, respectively. Let also the functions $q_2(t)$ and $\alpha(t)$ be the solutions of the differential equations~\eqref{q2dif} and \eqref{alpha2dif}, respectively. Finally, define the function $\kappa(t)$ according to equation~\eqref{kappa11}. Then equations
\begin{gather}
F\big(3^{-1/3}x, 3^{-2/3}t;\beta=6\big) = (\Psi(x,t))_{11}, \nonumber\\
 \Psi(x,t) = \kappa e^{\frac{x^3}{6} -\frac{xt}{2}}R(x,t)e^{-\frac{i\pi}{2}\sigma_3}
u^{-\frac{1}{2}\sigma_3} \Psi_{0}(x,t),\label{BVsol0}
\end{gather}
where
\begin{gather*}
R(x,t) = \begin{pmatrix} \dfrac{1+q_2}{2}x -\alpha&-1\vspace{1mm}\\
\dfrac{1-q^2_2}{4} &0\end{pmatrix}
\end{gather*}
define a solution $F(x,t;\beta=6)$ of the Bloemendal--Vir\'ag equation \eqref{bv1}. These formulae can be written as the following single equation
\begin{gather}
F\big(3^{-1/3}x, 3^{-2/3}t;\beta=6\big) \nonumber\\
\qquad{} = -i\kappa e^{\frac{x^3}{6} -\frac{xt}{2}}\left[ u^{-\frac{1}{2}}\left(\frac{1+q_2}{2}x -\alpha\right)\Psi_{011}(x,t) +u^{\frac{1}{2}}\Psi_{021}(x,t)\right].\label{BVsol}
\end{gather}
\end{Theorem}
Our goal now is to f\/ix solutions of the Painlev\'e II equation and of the equations~(\ref{q2dif}) and~(\ref{alpha2dif}) so that equation~(\ref{BVsol}) def\/ines the unique solution of (\ref{bv1}) which produces the $\beta=6$ Tracy--Widom distribution function. We remind that this solution is uniquely determined by the following boundary conditions
\begin{gather}\label{BVboundary1}
F(x,t;\beta) \to 1 , \qquad\text{as}\quad x, t \to +\infty,
\\
\label{BVboundary2}
F(x,t;\beta) \to 0,\qquad \text{as} \quad x \to -\infty, \quad t \leq t_0 < \infty.
\end{gather}
The $\beta = 6$ Tracy--Widom distribution function $F_{6}(t)$ is then given by the equation
\begin{gather*}
F_{6}(t) = \lim_{x\to +\infty}F(x, t;\beta=6).
\end{gather*}
Let us translate the boundary conditions (\ref{BVboundary1}), (\ref{BVboundary2}) to the relevant boundary conditions for the functions~$\Psi_0$, $q_2$ and $\alpha$. We start with $\Psi_0$.

The linear system
\begin{gather*}
\frac{d\Psi_0}{dx} = \hat{L}_0 \Psi_0
\end{gather*}
has six {\it canonical} solutions, $\Psi^{(k)}_0(x)$ which are characterized by the following asymptotic behavior in the complex $x$-plane (for all the details see, e.g.,~\cite{FIKN})
\begin{gather}\label{canon1}
\Psi^{(k)}_0(x) \sim \left(I + \frac{m_1}{x} + \cdots \right) e^{\left(\frac{x^3}{6} -\frac{xt}{2}\right)\sigma_3},
\qquad x \to \infty,\\
\frac{\pi}{2} +\frac{k-2}{3}\pi < \arg x < \frac{\pi}{2} +\frac{k}{3}\pi, \qquad k = 1, 2, \dots,7.\nonumber
\end{gather}
The second Painlev\'e function $u(t)$ can be recovered from the coef\/f\/icient~$m_1$ of the expansion~(\ref{canon1})
\begin{gather}\label{um1}
u = -m_{1,12} = m_{1,21}.
\end{gather}
The canonical solutions are entire functions of $x$ and they are related by {\it Stokes matrices}
\begin{gather}\label{Stokes}
\Psi^{(k+1)}_0(x)= \Psi^{(k)}_0(x)S^{(k)}_0, \qquad k = 1,2, \dots, 6.
\end{gather}
Also, one has
\begin{gather}\label{cyclic1}
\Psi^{(7)}_0(x)= \Psi^{(1)}_0(x).
\end{gather}
The Stokes matrices have the following structure:
\begin{gather}
S^{(1)}_0 = \begin{pmatrix} 1 &0\\ -is_1&1\end{pmatrix},\qquad
S^{(2)}_0 = \begin{pmatrix} 1 &is_2\\ 0&1\end{pmatrix}, \qquad
S^{(3)}_0 = \begin{pmatrix} 1 &0\\ -is_3&1\end{pmatrix},\nonumber\\
S^{(4)}_0 = \begin{pmatrix} 1 &-is_1\\ 0&1\end{pmatrix},\qquad
S^{(5)}_0 = \begin{pmatrix} 1 &0\\ is_2&1\end{pmatrix}, \qquad
S^{(6)}_0 = \begin{pmatrix} 1 &-is_3\\ 0&1\end{pmatrix}.\label {stokes2}
\end{gather}
where $s_1$, $s_2$ , $s_3$ can be any complex numbers subject the {\it cyclic relation}, which follows from~(\ref{cyclic1})
\begin{gather*}
s_1 -s_2 +s_3 + s_1s_2s_3 =0.
\end{gather*}
We shall restrict ourselves by considering only the real Painlev\'e functions $u$, which is equivalent to the additional restrictions on the Stokes parameters
\begin{gather*}
s_1 = \overline{s_3}, \qquad s_2 = \overline{s_2}.
\end{gather*}
Moreover, we shall be concerned with the Ablowitz--Segur family of the solutions which means the further restrictions
\begin{gather*}
s_2=0, \qquad s_1 = -ia = -s_3 , \qquad a\in {\mathbb R}, \qquad |a|\leq 1.
\end{gather*}
For all $a$, the solution $u(t)$ decays exponentially as $t \to +\infty$. In fact one has
\begin{gather}\label{as1}
u(t) = \frac{a}{2\sqrt{\pi}}t^{-\frac{1}{4}}e^{-\frac{2}{3}t^{\frac{3}{2}}} ( 1 + o(1) ),\qquad t \to +\infty.
\end{gather}
If $|a| <1$, then the solution $u(t)$ decays and oscillates as $t \to -\infty$, while if $|a| =1$ (the Hastings--McLeod solution), the function $u(t)$ grows as $|t|^{1/2}$
\begin{gather*}
u(t) = \pm\sqrt{-\frac{t}{2}} + O\left(\frac{1}{t}\right), \qquad t \to -\infty.
\end{gather*}
We shall show now that in order to formulae (\ref{BVsol0}) produce the solution of the Bloemendal--Vir\'ag equation (\ref{bv1}) satisfying the boundary conditions (\ref{BVboundary1}), (\ref{BVboundary2}) one has to choose the Hastings--McLeod Stokes data and to take $\Psi_0$ in~(\ref{BVsol0}) as
\begin{gather}\label{Psi0fix}
\Psi_0(x,t) = i\Psi^{(6)}_0(x,t)\sigma_1, \qquad \sigma_1 = \begin{pmatrix}0&1\\ 1&0\end{pmatrix}.
\end{gather}

Set
\begin{gather}\label{Ydef}
Y^{(6)}(x,t) := \Psi_{0}^{(6)}(x,t)e^{-\left(\frac{x^3}{6} -\frac{xt}{2}\right)\sigma_3}.
\end{gather}
We would have that, uniformly for all $t>0$
\begin{gather}\label{Yas1}
Y^{(6)}(x,t) \sim I + \frac{m_1(t)}{x} + \cdots, \qquad x\to +\infty
\end{gather}
(cf.~(\ref{canon1})). Substituting (\ref{Psi0fix})--(\ref{Yas1}) into (\ref{BVsol}) we have that
\begin{gather}
F\big(3^{-1/3}x,3^{-2/3}t;\beta=6\big) = \kappa e^{\frac{x^3}{6} -\frac{xt}{2}}\left[ u^{-\frac{1}{2}}\left(\frac{1+q_2}{2}x -\alpha\right)\Psi^{(6)}_{012}(x,t)+u^{\frac{1}{2}}\Psi^{(6)}_{022}(x,t)\right]\nonumber\\
\label{F22}
\hphantom{F\big(3^{-1/3}x,3^{-2/3}t;\beta=6\big)}{}
=\kappa u^{\frac{1}{2}}\left[u^{-1}\left(\frac{1+q_2}{2}x -\alpha\right)Y^{(6)}_{12}(x,t)
+Y^{(6)}_{22}(x,t)\right].
\end{gather}
From (\ref{F22}), in view of (\ref{Yas1}) and (\ref{um1}), we then get that
\begin{gather}\label{Fas22}
F\big(3^{-1/3}x,3^{-2/3}t;\beta=6\big) = \kappa u^{\frac{1}{2}}\left(\frac{1- q_2 }{2} +O\left(\frac{1}{x}\right)\right),\qquad x\to +\infty,
\end{gather}
uniformly for $t>0$. Assume now, that equations (\ref{q2dif}), (\ref{alpha2dif}) admit the solutions with the following behavior as $t \to +\infty$
\begin{gather}\label{assum0}
q_2(t) = -1 + o(1), \qquad \alpha = o(1) , \qquad t \to +\infty.
\end{gather}
Taking also into account the exponential decay (\ref{as1}) of the Hastings--McLeod solution $u(t)$ of the Painlev\'e equation (\ref{P2}), we conclude from~(\ref{kappa11}) that
\begin{gather*}
\frac{\kappa_t}{\kappa} \sim -\frac{1}{2}\frac{u_t}{u}, \qquad t \to +\infty.
\end{gather*}
Therefore, $\kappa(t)$ can be def\/ined in such a way that
\begin{gather*}
\kappa u^{\frac{1}{2}} \to 1, \qquad t \to +\infty.
\end{gather*}
This, together with estimate (\ref{Fas22}), implies the f\/irst boundary condition~(\ref{BVboundary1}) for the function $F(x,t;\beta)$.

To see what we have for the second boundary condition we f\/irst use the Stokes equations~(\ref{Stokes}) and the triviality of the Stokes matrix $S^{(2)}_0$ to rewrite $\Psi^{(6)}_0(x,t)$ as
\begin{gather}
\Psi^{(6)}_0(x,t) = \Psi^{(3)}_0(x,t) S^{(3)}_0 S^{(4)}_0
= \Psi^{(3)}_0(x,t)\begin{pmatrix}1&0\\ a&1\end{pmatrix}\begin{pmatrix}1&-a\\ 0&1\end{pmatrix}\nonumber\\
\hphantom{\Psi^{(6)}_0(x,t)}{}
= \Psi^{(3)}_0(x,t)\begin{pmatrix}1&-a\\ a&1-a^2\end{pmatrix}.\label{Psi63}
\end{gather}
Write (cf.~(\ref{Ydef}))
\begin{gather}\label{Ydef1}
Y^{(3)}(x,t) := \Psi_{0}^{(3)}(x,t)e^{-\left(\frac{x^3}{6} -\frac{xt}{2}\right)\sigma_3}.
\end{gather}
We would have that, this time, for every f\/inite $t$
\begin{gather}\label{Yas11}
Y^{(3)}(x,t) \sim I + \frac{m_1(t)}{x} + \cdots, \qquad x\to -\infty.
\end{gather}
Substituting (\ref{Psi0fix}), (\ref{Psi63}), (\ref{Ydef1}), and (\ref{Yas11}) into (\ref{BVsol}) we will obtain an alternative to (\ref{F22}) representation for the function $F(x,t;\beta)$
\begin{gather}
F\big(3^{-1/3}x,3^{-2/3}t;\beta=6\big) = (1-a^2)\kappa u^{\frac{1}{2}}\left[u^{-1}\left(\frac{1+q_2}{2}x -\alpha\right)Y_{12}^{(3)}(x,t) + Y^{(3)}_{22}(x,t)\right]\nonumber\\
\label{F222}
\qquad{}
-a\kappa u^{\frac{1}{2}}e^{\frac{x^3}{3} - xt}\left[u^{-1}\left(\frac{1+q_2}{2}x -\alpha\right)Y_{11}^{(3)}(x,t)
+Y^{(3)}_{21} (x,t) \right].
\end{gather}
Taking into account that
 \begin{gather*}
 e^{\frac{x^3}{3} -xt} \to 0, \qquad x \to -\infty, \qquad t \leq t_0 < \infty,
 \end{gather*}
 and that $a^2 = 1$ for the Hastings--McLeod solution, we arrive at the second boundary condition~(\ref{BVboundary2}) for the function~$F(x,t;\beta)$.

Our analysis can be summarized as the following proposition.
\begin{Proposition}\sloppy Let $u(t)$ be the Hastings--McLeod solution of the second Painlev\'e equa\-tion~\eqref{P2} and $\Psi^{(6)}$ be the canonical solution of the corresponding isomonodromy linear problem. Suppose that equations~\eqref{q2dif}, \eqref{alpha2dif} have solutions which are smooth for all real $t$ and satisfy condi\-tions~\eqref{assum0} at $t=+\infty$. Then formulae \eqref{BVsol}, \eqref{Psi0fix} define the $($unique$)$ solution of the Bloemendal--Vir\'ag equation~\eqref{bv1} satisfying the boundary conditions \eqref{BVboundary1}, \eqref{BVboundary2}. This, in turn, yields the following formula for the $\beta=6$ Tracy--Widom distribution function
\begin{gather*}
 F_{6}\big(3^{-2/3}t\big) = \frac{1-q_2}{2}
\exp\left(\frac{1}{3}\int_{t}^{\infty}\omega(s)ds +\frac{2}{3}\int_{t}^{\infty}\alpha(s)ds
 -\frac{1}{3}\int_{t}^{\infty}\frac{u_s(s)}{u(s)}(1+q_2(s))ds\right),
 \end{gather*}
 which, taking into account \eqref{alphaq2}, can be also written as
 \begin{gather}\label{TWanswer2}
 F_{6}\big(3^{-2/3}t\big) = \frac{(q_2-1)}{2q_2}
 \exp\left(\frac{1}{3}\int_{t}^{\infty}\omega(s)ds
 -\frac{2}{3}\int_{t}^{\infty}\frac{u_s(s)}{u(s)}\frac{1+q_2(s)}{q_2(s)}ds\right).
 \end{gather}
 \end{Proposition}

Formula (\ref{TWanswer2}) is equivalent, though not identical, to the original formula~(1.13) of~\cite{Rumanov}.

\section[Asymptotics of $F_{6}(t)$ as $t\to-\infty$]{Asymptotics of $\boldsymbol{F_{6}(t)}$ as $\boldsymbol{t\to-\infty}$} \label{assymp}

In this section we show that, at least on the formal level, equation (\ref{TWanswer2}) can be used to evaluate the asymptotics of $F_{6}(t)$ as $ t \to -\infty$. Similar fact involving the original Rumanov's formula has already been demonstrated in~\cite{Rumanov}.

To make step towards formula (\ref{con1}) we have to f\/ind asymptotics of all the integrands in~(\ref{TWanswer2}). The f\/irst ingredient of our computation is the well-known formal asymptotic expansion of the Hastings--McLeod solution
\begin{gather*}
u(t)=\sqrt{-\frac{t}{2}} \left(1 -\frac{1}{8}(-t)^{-3} -\frac{73}{128}(-t)^{-6} -\frac{10657}{1024}(-t)^{-9}\right.\\
\left.\hphantom{u(t)=}{}
-\frac{13912277}{32768}(-t)^{-12} -\frac{8045883943}{262144}(-t)^{-15} -\frac{14518451390349}{4194304}(-t)^{-18}+\cdots \right),
\end{gather*}
and the expansion of its logarithmic derivative
\begin{gather}
\frac{u_t(t)}{u(t)}= -\frac{1}{2(-t)} -\frac{3}{8}(-t)^{-4} -\frac{111}{32}(-t)^{-7} -\frac{1509}{16}(-t)^{-10}\nonumber\\
\hphantom{\frac{u_t(t)}{u(t)}=}{}
-\frac{2617599}{512}(-t)^{-13} -\frac{944695983}{2048}(-t)^{-16} -\frac{127756233309}{2048}(-t)^{-19} +\cdots.\label{dlogHM_as}
\end{gather}
The above expansions immediately yield the formal expansion of the Hamiltonian function $\omega$ def\/ined in (\ref{r011})
\begin{gather}
\omega(t)=-\frac{1}{4}(-t)^2 -\frac{1}{8}(-t)^{-1} -\frac{9}{64}(-t)^{-4} -\frac{189}{128}(-t)^{-7} -\frac{21663}{512}(-t)^{-10}\nonumber\\
\hphantom{\omega(t)=}{}
-\frac{4825971}{2048}(-t)^{-13} -\frac{3540311739}{16384}(-t)^{-16} -\frac{241980297111}{8192}(-t)^{-19} +\cdots.\label{omega_as}
\end{gather}

In contrast, the formal expansion of $q_2(t)$ as $t\to-\infty$ is much less straightforward and requires relatively signif\/icant ef\/forts.
With this aim, we utilize the Rumanov's linearization (\ref{mu+})--(\ref{nu}) and then apply~(\ref{q2mu}).

The corresponding coef\/f\/icient matrix ${\mathcal M}(t)$ of the vector equation for $\vec\mu:=(\mu_+,\mu_-,\nu)^T$, i.e., $\vec\mu_t={\mathcal M}\vec\mu$,
\begin{gather*}
{\mathcal M}(t)=
\begin{pmatrix}
\dfrac{2}{3}\dfrac{u_t}{u}&0&-\dfrac{1}{3}\vspace{1mm}\\
0&-\dfrac{2}{3}\dfrac{u_t}{u}&\dfrac{1}{3}\vspace{1mm}\\
\dfrac{2}{3}\dfrac{\omega}{u^2}& \dfrac{2}{3}u^2&0
\end{pmatrix}=
\begin{pmatrix}
-\dfrac{1}{3(-t)}+{\mathcal O}(t^{-4})&0&-\dfrac{1}{3}\vspace{1mm}\\
0&\dfrac{1}{3(-t)}+{\mathcal O}(t^{-4})&\dfrac{1}{3}\vspace{1mm}\\
-\dfrac{(-t)}{3}+{\mathcal O}\big(t^{-2}\big)& \dfrac{(-t)}{3}+{\mathcal O}\big(t^{-2}\big)&0
\end{pmatrix}
\end{gather*}
degenerates in the leading order at inf\/inity. Thus we f\/irst apply the shearing gauge transformation
\begin{gather*}
\vec\mu=R_0\vec y,\qquad
R_0=
\begin{pmatrix}
(-t)^{-1/4}&0&0\\
0&(-t)^{-1/4}&0\\
0&0&(-t)^{1/4}\\
\end{pmatrix}
\begin{pmatrix}
1&1&1\\
1&-1&-1\\
0&\sqrt2&-\sqrt2
\end{pmatrix},
\end{gather*}
that diagonalizes in the leading order the coef\/f\/icient matrix
\begin{gather*}
\vec y_t={\mathcal N}\vec y,\qquad {\mathcal N}=R^{-1}{\mathcal M}R-R^{-1}R_t\\
\hphantom{\vec y_t={\mathcal N}\vec y,\qquad {\mathcal N}}{}=
\begin{pmatrix}
-\frac{1}{4(-t)}+{\mathcal O}\big(t^{-20}\big)&
-\frac{1}{3(-t)}+{\mathcal O}\big(t^{-7}\big)&
-\frac{1}{3(-t)}+{\mathcal O}\big(t^{-4}\big)\\
-\frac{1}{6(-t)}+{\mathcal O}\big(t^{-5/2}\big)&
-\frac{\sqrt2}{3}\sqrt{-t}+{\mathcal O}\big(t^{-5/2}\big)&
-\frac{1}{4(-t)}+{\mathcal O}\big(t^{-5/2}\big)\\
-\frac{1}{6(-t)}+{\mathcal O}\big(t^{-5/2}\big)&
-\frac{1}{4(-t)}+{\mathcal O}\big(t^{-5/2}\big)&
\frac{\sqrt2}{3}\sqrt{-t}+{\mathcal O}\big(t^{-5/2}\big)
\end{pmatrix},
\end{gather*}
that enables us to construct ef\/fectively the formal asymptotic expansion of $\vec y(t)$ as $t\to-\infty$. Three independent vector solutions to the above linear ODE form a matrix $Y(t)$
\begin{gather*}
Y(t)=\left(I+\sum_{k=1}^{\infty}Y_k(-t)^{-3k/2}\right)
\begin{pmatrix}
(-t)^{1/4}&0&0\\
0&e^{\frac{2\sqrt2}{9}(-t)^{3/2}}&0\\
0&0&e^{-\frac{2\sqrt2}{9}(-t)^{3/2}}
\end{pmatrix},
\end{gather*}
where $Y_j$ are independent from $t$ matrix coef\/f\/icients
\begin{gather*}
Y_1= \begin{pmatrix}
0&\dfrac{1}{\sqrt2}&-\dfrac{1}{\sqrt2}\vspace{1mm}\\
-\dfrac{1}{2\sqrt2}&-\dfrac{11}{48\sqrt2}&-\dfrac{3}{8\sqrt2}\vspace{1mm}\\
\dfrac{1}{2\sqrt2}&\dfrac{3}{8\sqrt2}& \dfrac{11}{48\sqrt2}
\end{pmatrix},
\qquad
Y_2= \begin{pmatrix}
-\dfrac{1}{9}&\dfrac{259}{96}&\dfrac{259}{96}\vspace{1mm}\\
\dfrac{1}{2}&-\dfrac{665}{3072}&\dfrac{113}{256}\vspace{1mm}\\
\dfrac{1}{2}&\dfrac{113}{256}&-\dfrac{665}{3072}
\end{pmatrix},\\
Y_3= \begin{pmatrix}
0&\dfrac{83803}{3072\sqrt2}&-\dfrac{83803}{3072\sqrt2}\vspace{1mm}\\
-\dfrac{347}{72\sqrt2}&-\dfrac{1733015}{1327104\sqrt2}&-\dfrac{60101}{24576\sqrt2}\vspace{1mm}\\
\dfrac{347}{72\sqrt2}&\dfrac{60101}{24576\sqrt2} &\dfrac{1733015}{1327104\sqrt2}
\end{pmatrix}
,\qquad\dots.
\end{gather*}
The gauge transformation $R_0$ does not mix the vector columns, thus we have three possible solutions $q_2(t)$ distinguished by their asymptotic behavior as $t\to-\infty$ according to which basic vector solution $\vec y$ dominates in the relevant combination
\begin{gather}
q_2(t)= \begin{cases}
(-t)^3-\dfrac{225}{4} +{\mathcal O}\big(t^{-3}\big),\qquad \vec y\sim(-t)^{1/4}, \vspace{1mm}\\
\dfrac{1}{\sqrt2}(-t)^{-3/2} +\dfrac{21}{8}(-t)^{-3} +\dfrac{1707}{64\sqrt2}(-t)^{-9/2} +{\mathcal O}\big(t^{-6}\big),\qquad
\vec y\sim e^{\frac{2\sqrt2}{9}(-t)^{3/2}}, \vspace{1mm}\\
-\dfrac{1}{\sqrt2}(-t)^{-3/2} +\dfrac{21}{8}(-t)^{-3} -\dfrac{1707}{64\sqrt2}(-t)^{-9/2} +{\mathcal O}\big(t^{-6}\big),\qquad
\vec y\sim e^{-\frac{2\sqrt2}{9}(-t)^{3/2}}.
\end{cases}\hspace{-10mm}\label{q2_m8_as}
\end{gather}
Finally, using (\ref{dlogHM_as}), (\ref{omega_as}) and the second choice in the expansion (\ref{q2_m8_as}) (this means that in an exact description of $q_2(t)$, the dominant vector $\vec\mu$ is presented) we f\/ind that
\begin{gather*}
\frac{d}{dt}\log F_{6}\big(3^{-2/3}t\big)=\frac{1}{12}t^2-\frac{\sqrt{2}}{3}(-t)^{1/2}+\frac{1}{24\,t}+O\big(|t|^{-\frac{5}{2}}\big)\qquad \mbox{as \ \ $t\to-\infty$},
\end{gather*}
so that by the scaling $t\to 3^{2/3}t$ one obtains
\begin{gather*}
\frac{d}{dt}\log F_{6}(t)=\frac{3}{4}t^2-\sqrt{2}(-t)^{1/2}+\frac{1}{24 t}+O\big(|t|^{-\frac{5}{2}}\big)\qquad \mbox{as \ \ $t\to-\infty$},
\end{gather*}
which after integration gives
\begin{gather}\label{FTW6}
\log F_{6}(t)=-\frac{1}{4}|t|^3+\frac{2\sqrt{2}}{3}|t|^{3/2}+\frac{1}{24}\log |t|+c_0+O\big(|t|^{-\frac{3}{2}}\big)\qquad \mbox{as \ \ $t\to-\infty$},
\end{gather}
for some constant $c_0$. The above expression coincides with~(\ref{con1}) for $\beta=6$. In our derivation the quantity $\chi$ is an undetermined constant.

\section{Open questions}

In this f\/inal section we highlight the two principal open questions in our version of Rumanov's scheme which are needed to be answered in order to make the method complete.
\begin{enumerate}\itemsep=0pt
\item Prove that indeed the system (\ref{q2dif}), (\ref{alpha2dif}) has global smooth solution satisfying Cauchy conditions (\ref{assum0})
 at $t=+\infty$.

 \item Assuming that the previous problem has been solved, establish that the solution $q_2(t)$ has the power expansion as $t \to -\infty$ which generates via equation~(\ref{TWanswer2}) the asymptotics for $F_{\beta=6}(t)$ obtained in~\cite{BEMN}. Here, the main challenge is to prove that the solution with the Cauchy data $q_2 (+\infty) = -1$ and $q_{2t}(+\infty) =0$ has indeed the needed power series expansion at $t = -\infty$. This is a {\it connection problem}, and we strongly believe that it would be very dif\/f\/icult to solve it without establishing the Lax-integrability of equations (\ref{q2dif}), (\ref{alpha2dif}). It is important to emphasize that formally, the needed expansions of $q_2(t)$ at $t=+\infty$ and $t=-\infty$ could be found by a direct perturbation analysis of equations~(\ref{q2dif}),~(\ref{alpha2dif}). For that, integrability is really not needed; indeed, this has already been done in~\cite{Rumanov} and demonstrated in Section \ref{assymp} of this paper. The real issue is to prove that these expansions are expansions of the {\it same} solution.
\end{enumerate}

Both problems indicated above, will be solved if, for instance, one succeeds in the reduction of the linear version of equations (\ref{q2dif}), (\ref{alpha2dif}), i.e., of the equations (\ref{mu+})--(\ref{nu}) to one of the known Lax pairs for the second Painlev\'e equation as it is discussed at the end of Section~\ref{S5}. We intend to address all these questions in our next publication as well as the issue of the extension of these results to the all even values of~$\beta$.

Before concluding this paper we want to make some extra comments on the linearizability of equations (\ref{q2dif}), (\ref{alpha2dif}) and on the relevance of this fact to our principal goal, i.e., to the proof of Conjecture \ref{bemncon1}. The fact that these equations, as well as equation~(\ref{q2eq3}) and Rumanov's equation~(\ref{etaeq}), are linearizable, is, of course, very important, but in itself is not enough to solve the above mentioned connection problem and hence to prove Conjecture~\ref{bemncon1}. Indeed, usually, in order to solve connection problems\footnote{By solving a connection problem we mean to solve it {\it explicitly}, that is in terms of elementary or known special functions, i.e., exactly in the form which we need solution of our problem in order to prove~(\ref{FTW6}) and Conjecture~\ref{bemncon1}.} for a linear equation with rational coef\/f\/icients, one needs to have some additional information about its solutions. Most often this addition information is given in the form of contour integral representation which is available through the Laplace's method and only for very special linear equations, i.e., for hypergeometric equation and its degenerations. In the nonlinear case, or in the case of linear equations with meromorphic coef\/f\/icients (as it is the case with equations (\ref{mu+})--(\ref{nu})), Laplace's method is replaced by the Riemann--Hilbert method and the contour integral representation is replaced by the Riemann--Hilbert representation. The Riemann--Hilbert method is as ef\/fective for solving connection problems for nonlinear equations as Laplace's method in the linear case (see, e.g.,~\cite{FIKN}). However, for the applicability of the Riemann--Hilbert method one needs Lax pairs. Hence our desire to have a Lax-pair formulation either for equations (\ref{q2dif}), (\ref{alpha2dif}) themselves or for their linear version (\ref{mu+})--(\ref{nu}).

We also want to mention one more interesting observation. Linearizability of the second order dif\/ferential equations (\ref{q2eq3}) and (\ref{etaeq}) mean that they possess the Painlev\'e property{\footnote{The solutions of these equations do not have movable branch points; all their possible branch points are at the poles of the coef\/f\/icients of the equations. Indeed, in~\cite{Rumanov} all the relevant exponents have been calculated and the absence of logarithmic terms has been established.}} and hence must be equivalent to one of the 50 canonical equations from the Gambier list, see~\cite{Ince}. Let us take Rumanov's equation~(\ref{etaeq}) and make the following substitutions,
\begin{gather*}
\eta(t) = \lambda(t)W(z) + \zeta(t), \qquad z = \xi(t),
\end{gather*}
where the local analytic change-of-variable functions $\lambda(t)$, $\zeta(t)$, and $\xi(t)$ are def\/ined through the equations,
\begin{gather}\label{zeta0}
9\zeta_{tt} + 9\zeta\zeta_t +\zeta^3 + P(t)\zeta + Q(t) = 0,\\
9\lambda_{tt} +9\zeta \lambda_t + \bigl(9\zeta_t +3\zeta^3 + P(t)\bigr)\lambda = 0,\nonumber
\end{gather}
and
\begin{gather*}
\xi_t = \frac{1}{3}\lambda.
\end{gather*}
Then, equation (\ref{etaeq}) transforms to the equation
\begin{gather}\label{ince}
\frac{d^2W}{dz^2}= -3W\frac{dW}{dz} - W^3 + v(z) \left\{\frac{dW}{dz} + W^2\right\},
\end{gather}
where
\begin{gather*}
v = -\frac{3\zeta\lambda +9\lambda_t}{\lambda^2}.
\end{gather*}
This is equation \# VI from the list given in the Ince monograph~\cite{Ince}. This equation is linearized by the substitution, $W = -\frac{d\ln F}{dz}$. Moreover, the equation on~$F$ is
\begin{gather*}
F_{zzz} = v F_{zz},
\end{gather*}
and hence is solvable in quadratures. Unfortunately, one of the change~-- of variables equations~-- equation~(\ref{zeta0}), is again Rumanov's equation (\ref{etaeq}). Hence, though theoretically important, the reduction of~(\ref{etaeq}) to~(\ref{ince}) does not immediately help in the achievement of our main goal, i.e., to prove~(\ref{FTW6}).

\subsection*{Acknowledgements}
A.~Its and T.~Grava acknowledge the support of the Leverhulme Trust visiting Professorship grant VP2-2014-034. A.~Its acknowledges the support by the NSF grant DMS-1361856 and by the SPbGU grant N~11.38.215.2014. A.~Kapaev acknowledges the support by the SPbGU grant N~11.38.215.2014. F.~Mezzadri was partially supported by the EPSRC grant no.~EP/L010305/1. T.~Grava acknowledges the support by the Leverhulme Trust Research Fellowship RF-2015-442 from UK and PRIN Grant ``Geometric and analytic theory of Hamiltonian systems in f\/inite and inf\/inite dimensions'' of Italian Ministry of Universities and Researches.

\pdfbookmark[1]{References}{ref}
\LastPageEnding

\end{document}